\documentstyle[aps,twocolumn,prl,psfig]{revtex}

\begin{document}

\title{Magnetic Field resulting from non-linear electrical transport in single crystals 
of  charge-ordered Pr$_{0.63}$ Ca$_{0.37}$ MnO$_{3}$} 

\author{Ayan Guha $^*$}
\address{Department of Physics, Indian Institute of Science, Bangalore 560 012, India}
\author{N.Khare and A.K.Raychaudhuri}
\address{National Physical Laboratory, Dr. K.S. Krishnan Marg, New Delhi-110012, India}
\author{C.N.R Rao}
\address{$^*$CSIR Center of Excellence in Chemistry, Jawaharlal Nehru Center for Advanced 
Scientific Research, \mbox{Jakkur P.O., Bangalore 560 064}, India}
                  

\twocolumn[\hsize\textwidth\columnwidth\hsize\csname 
@twocolumnfalse\endcsname

\maketitle
\begin{abstract}

In this letter we report that the current induced destabilization of  
the charge ordered (CO) state in a rare-earth manganite gives rise to 
regions with ferromagnetic correlation. We did this experiment by 
measurement of the I-V curves in single crystal of the CO system
 Pr$_{0.63}$Ca$_{0.37}$MnO$_{3}$ and simultanously measuring the 
magnetization of the current carrying conductor using a high T$_c$ SQUID
working at T = 77K. We have found that the current induced            
destabilization of the CO state leads to a regime of 
negative differential resistance which leads to a small enhancement 
of the magnetization of the sample, indicating ferromagnetically aligned moments.

\end{abstract}
\pacs {}

]

Electrical transport in rare-earth manganites have attracted considerable current interest 
because of a number of novel properties, like Colossal magnetoresistance (CMR) and 
Charge-ordering(CO)~\cite{Tokura1,Raorev}. These manganites belong to the ABO$_{3}$ type 
perovskite oxides and have a general chemical formula Re$_{(1-x)}$ A$_{x}$ MnO$_{3}$ where 
Re is the rare-earth such as La,Nd and Pr and A is an alkaline-earth element such
 as Ca,Sr and Ba.  These manganites, depending on the size of the average A
site cationic radius, can charge order, in particular, when x=1/2,2/3,4/5 etc.The formation 
of the CO state can also occur for other incommensurate values of the carrier concentration.

\vspace{0.2cm}
The CO state is strongly destabilized by different types of perturbations which include magnetic 
field~\cite{Tokura1}, electric field~\cite{Tokura2,Ponna1} and optical radiation~\cite{Ogawa1,Fiebig1}. 
Application of an applied magnetic field of sufficient magnitude can lead to a 
collapse of the CO gap, $\Delta_{co}$, at the Fermi level and the melting of the charge-ordered 
insulating (COI) state to a ferromagnetic metallic (FMM) state~\cite{Okimoto98,Biswas992}. Laser 
radiation creates conducting filaments which at low temperatures lead to non-linear 
transport~\cite{Ogawa1}. Application of an electric field beyond a threshold value also gives rise 
to non-linear conduction accompanied by  a broad band noise of substantial magnitude~\cite{Ayan1}. 
. A topic of considerable current interest is  what causes the 
destabilization of 
the CO state and whether the underlying mechanism is the same for all the perturbations.

\vspace{0.2cm}
An important issue associated with the formation of the COI state is that of spin ordering. 
For T $>$ T$_{CO}$ (i.e, in the PM phase), the dominant spin correlation is FM which grows on 
cooling~\cite{Kajimoto1}. At T $<$ T$_{CO}$, the FM spin correlations decrease till AFM ordering 
sets in at T$_{N}$. The COI state is generally stabilized by AFM spin correlations. The metallic  
state, on the other hand, is stabilized by FM spin correlations. The presence of ferromagnetic 
interaction inhibits the formation of the COI state. In other words, the destabilization of the 
COI state by any interaction or external stimulus is expected to suppress AFM spin correlations 
and promote FM spin correlations. One can therefore surmise that when COI state is destabilized 
by whatever means, there should be a magnetic signature of the transition in terms of the 
enhanced magnetic moment.

In this letter, we have tested the above hypothesis in single crystals of charge ordered 
Pr$_{0.63}$ Ca$_{0.37}$ MnO$_{3}$ by simultaneous measurement of I-V characteristics and 
detection of the magnetic moment by a high T$_{c}$ SQUID.

\vspace{0.2cm}
Single crystals of Pr$_{0.63}$ Ca$_{0.37}$ MnO$_{3}$ were grown by float zone technique in an 
image furnace. The crystal used for our experiment was cut to a size of 4 x 2 x 0.3 mm$^3$ and 
current and voltage probes were attached to it with Ag-In alloy with a mean separation of the 
probes $ \approx$ 0.25mm. The experiment was done at T=77K by dipping the sample in liquid 
nitrogen. The non-linear I-V characteristics were taken by  biasing the sample with a constant 
current supply. The high T$_C$ SQUID~\cite{Neeraj1} used for the detection of the magnetic 
moment was based on BSCCO film operating at T $\approx$ 77K which is capable of detecting a 
field weaker than 10$^{-10}$ T.

\vspace{0.2cm}
The schematic of the RF SQUID set-up used to detect the small magnetic field associated with the 
destabilization of the COI
state is shown in figure~1. The high T$_C$ SQUID is based on natural grain boundary junction in 
BSCCO film (T$_C$ $ \approx $ 108K) deposited on a single crystal SrTiO$_3$ substrate by screen 
printing technique. The SQUID geometry consists of a hole of diameter 300$\mu$m and a microbridge 
of width 80$\mu$m. RF-biasing of the SQUID is applied through a tank circuit whose resonance 
frequency and Q are 17.5MHz and 50 respectively. The inductance of the tank circuit consists of 
three coils of diameter approxiately 1mm : two of 25 turns(\# 38 SWG copper) and one of 8 turns. 
The 8 turn coil is mounted concentrically on the hole on the surface of the BSCCO film. Three 
concentric cylinders of $\mu$ metal are used to shield the SQUID and the sample from the external 
magnetic field. The tank circuit was driven at V$_{rf}$ $\approx$ 750$\mu$V. Phase sensitive 
detection using a lock-in amplifier (LIA) was implemented by an audio frequency (f $\approx$ 
300Hz) field modulation ($\Delta$H$_m$/4 $\approx$ 1.6x10$^{-8}$T.
The field modulation $\Delta$H$_m$ corresponds to one flux quanta ($\Phi$$_o$). The transfer 
function of the SQUID field detector is 1.1X10$^{-10}$T/$\mu$V at the LIA. Thus from the LIA 
out put we can calulate the field generated by the sample.

\vspace{0.2cm}
Our detection was done without any biasing dc magnetic field and we didnot use a flux-locked 
loop. Given the small flux change seen in the experiment ( $\approx  \Phi_{o}$ ) it was a less 
noisy option compared to the flux locked loop detection mode. The flux noise S$_{\phi}$ of the 
detection circuit down to 1Hz had a spectral power of 10$^{-6}$ $\Phi_{o}^{2}$ /Hz which is much 
less than the magnetic signal detected in our experiment. The sample with the leads attached for 
taking the I-V measurements was put on top of the tank circuit. Thus high detectibility of the 
magnetic signal was ensured because of the close proximity of the SQUID detector with the sample 
and we can detect the magnetic signal simultaneously while taking the I-V data. It is the use of 
the high T$_{c}$ SQUID that enabled us to do the experiment at a elevated temperature of 77K, can 
be considered as the novelty of the experiment. The BSCCO SQUID used by us could be used for 
operation upto 100K.

\vspace{0.2cm}
In the inset of figure~ 2 we show the resistivity($\rho$) as a function of T down to 50K 
measured with a very small current of less than 0.01 mA which ensures that the COI state is 
not destabilized. The onset of the COI state at T$_{co}$ $\approx$ 240K is clearly identifiable. 
The AFM order sets in at a much lower temperature and from magnetic susceptibility measurements 
the T$_N$ is found to be $\approx$ 175K. The I-V curves recorded at few temperatures above and 
below T$_{co}$ show onset of non-linear conduction at all T $<$ T$_{co}$. The non-linearity 
increases as T is lowered and below $\approx $ 170K one 
sees the onset of negative differential resistance(NDR). The onset of the NDR regime in the 
I-V curve and its behavior in an applied magnetic field have been a subject of a detailed 
investigation and is being published elsewhere ~\cite{Ayan2}. Given the limited scope of 
this letter we focus only on one aspect namely
the magnetic signature of the NDR regime in the I-V curve. The 
current-induced creation of NDR is a reproducible effect and shows no change on repeated current 
cycling from positive to negative swing of bias current. 
The creation of NDR is also not due to heating as verified by directly measuring the sample 
temperature. The sample was directly dipped in liquid nitrogen. At the highest power dissipation 
the sample temperature was not more than 3K from the bath. Also, significant heating by the sample 
will make the SQUID output drop because the SQUID is in intimate thermal contact with the sample. 
The NDR region shows V $\propto$ I$^{-n}$ (1$>$n$>$0). 
 
\vspace{0.2cm}
In figure~ 3 we show the  results of SQUID measurements along with I-V curve at 77K. The output 
of the lockin amplifier (LIA) is plotted as function of the current flowing through the sample. 
The output of the LIA is $\propto$ the magnetization change ($\delta m$) of the sample created 
by passing the current through it. Our experimental arrangement allowed us to measure the I-V 
curve and the magnetic measurements simultaneously. We increased the bias current(I) 
in small steps 
and covered the range upto 15mA in 10$^4$ secs and waited for 20 sec approximately for 
stabilization after each current step before the lock-in reading and the voltage across the 
voltage leads were recorded. This method ensured that there is no effects from a changing 
measuring current. We find that as the sample enters the NDR regime at a current I $ \approx$ 
3mA, there is a clear rise in the signal from the LIA indicating a rise in the magnetization, 
$\delta m$, of the sample. $\delta m$ increases as the current I is increased although
a jump in $\delta m$ occurs when I $ \approx $ 3mA where the NDR regime sets in. For 5mA $<$ 
I $<$ 10mA, the LIA signal shows a plateau and increases further when I is increased beyond 
10mA. Our experiment thus establishes that the current-induced destabilization of the COI order 
leads to an enhancement of the magnetization (albeit small) indicating onset of ferromagnetic 
correlations.

\vspace{0.2cm}
It has been found recently that the non-linear conduction in the COI state is associated with 
a large voltage fluctuation~\cite{Ayan1}. We find that such a large voltage fluctuation does 
indeed appear in our system also at the onset of non-linear conduction. We show the magnitude 
of the voltage fluctuation $<$ $\delta$V$^{2}$ $>$/V$^{2}$ as a function of the applied current 
bias I in the inset of figure~4. Interestingly, we find that even this fluctuation has a magnetic 
signature. In figure~4, we show the
measured flux noise of the SQUID. The flux noise was measured by measuring the voltage noise of 
the LIA output by a spectrum analyzer. The flux noise was then calculated using the relation 
\mbox{S$_\phi$ = S$_v$ /($\delta$ v / $\delta\phi$)$^2$,} where S$_v$ is the spectral noise power of 
the voltage putput of the LIA and
$\delta$v / $\delta\phi$ is the transfer function of the SQUID. The flux noise below 1Hz has a 
1/f dependence and above
1Hz is a broad noise with superimposed peaks. At I = 0 mA the flux noise \mbox{S$_{\phi}$(I=0) 
$ \approx $10$^{-6}$ $\Phi_{o}^{2}$ /Hz} and at I = 8mA when the I-V curve is in the NDR regime, 
there is a clear rise in the flux noise and \mbox{S$_{\phi}$ (I = 8mA) $ \approx $ 
4x10$^{-6}$ $\Phi_{o}^{2}$/Hz.} This extra noise is from the sample. It appears that on passage 
of the current when the NDR regime (likely with a metallic filament) sets in, the resulting 
stage has strong spin fluctuation. 

\vspace{0.2cm}
The signal in the SQUID coil does not arise from the magnetic field of the current flowing 
through the sample. This has been directly tested by passing the same current through a dummy 
resistor instead of the sample. (Though the magnitude of the field produced by a direct measuring 
current at the SQUID is \mbox{$ \approx $ 0.2$\mu$T/mA,} the field at the SQUID is parallel to its 
plane and has negigible component perpendicular to it.)

\vspace{0.2cm}
We propose a simple scenario to explain the observations.  The onset of the NDR region in the 
I-V curve beyond certain value of I is due to appearance of the metallic filaments which carry 
most of the current. This decreases the voltage across the sample. On passing more current, the 
volume fraction of the filament increases leading to a further decrease of voltage and the NDR 
regime is sustained. The SQUID observations show that the filaments so formed have 
ferromagnetically aligned Mn moments which enhance the magnetization. We may consider the 
filaments as made up of the FMM phase. NDR is also seen in the I-V characteristics of 
single crystals of the same system on application of laser light. Direct observation using 
reflection meaurements show that in this region there is formation of "metallic" 
filament~\cite{Fiebig1}. Interestingly, a small jump in magnetization was also seen in laser 
induced melting. 
 
\vspace{0.2cm}
The output signal at the LIA which has been directly calibrated by applying a known dc signal in a 
field coil can be used to estimate magnetic moments generated on the sample. The transfer function of the SQUID field detector is $ \approx$ 1.1x10$^{-10}$T / $\mu$V measured at the LIA. At the highest value of sample current I, the field created by the sample is H$_{sample}$ $ \approx$ 35 nT. Knowing the approximate geometry of the detection scheme and assuming that the sample creates a dipolar field, we can then estimate the total moment formed on the
sample $\mu_{sample}$ $\approx$ 5x10$^{-15}$ Wb-m $ \approx$ 4x10$^{14}$ $\mu_B$.
We make a simple assumption that this moment arises from a ferromagnetic region with all the Mn$^{3+}$ and Mn$^{4+}$ aligned, the estimated volume of the region will then be V$_{fm}$ $\approx$ 6x10$^{12}$ nm$^3$. (This has 
been estimated from the lattice constant and saturation moment of 3.8$\mu_B$/Mn ion). This is a lower limit on V$_{fm}$. If we assume that the magnetic moment resides on the filaments of FMM region of volume V$_{fm}$ , as stated before and the filament has to span a length between the current pads (250$\mu$m) we can put a lower limit on its cross-section $\approx$ 24 $\mu$m$^2$. These simple estimates are based on assumptions which are not too rigorous nevertheless they set the scale of magnetic moment $\mu_{sample}$ and the V$_{fm}$ created by the application of the current.The important issue is that 
at the onset of NDR a small but finite magnetic moment appears on the  sample.

\vspace{0.2cm}
To conclude, we find that in charge-ordered Pr$_{0.63}$ Ca$_{0.37}$ MnO$_{3}$, a current can 
destabilize the CO state giving rise to a region of negative differential resistance which sets in beyond a 
current threshold. The appearance of the NDR region gives rise to a small increase in the magnetization 
suggesting that the NDR region arises from filaments of the FMM phase formed by the current.

\newpage
{\centerline {FIGURE CAPTIONS}}            

 FIG.~1.Schematic of the high T$_c$ SQUID setup (working at T=77K) used for measuring 
small changes in magnetization in the current carrying sample. 
 
FIG~2.The I-V curves showing nonlinear conduction and negative 
differential resistance. 
The resistivity $\rho$ of the sample is shown in the inset. The charge ordering temperature 
T$_{CO}$ is marked. 

FIG.~3.  The SQUID signal( output of the lock-in amplifier) as fuunction of current through 
the sample. The I-V curve  
is measured simultaneously.

FIG.~4. Measured flux noise seen in the SQUID for I = 0 and I=8mA. The inset shows the 
voltage noise in the sample simultaneously measured.

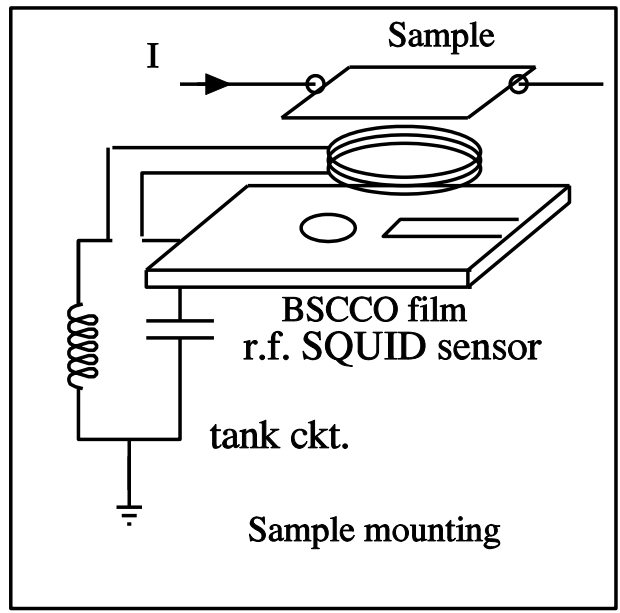
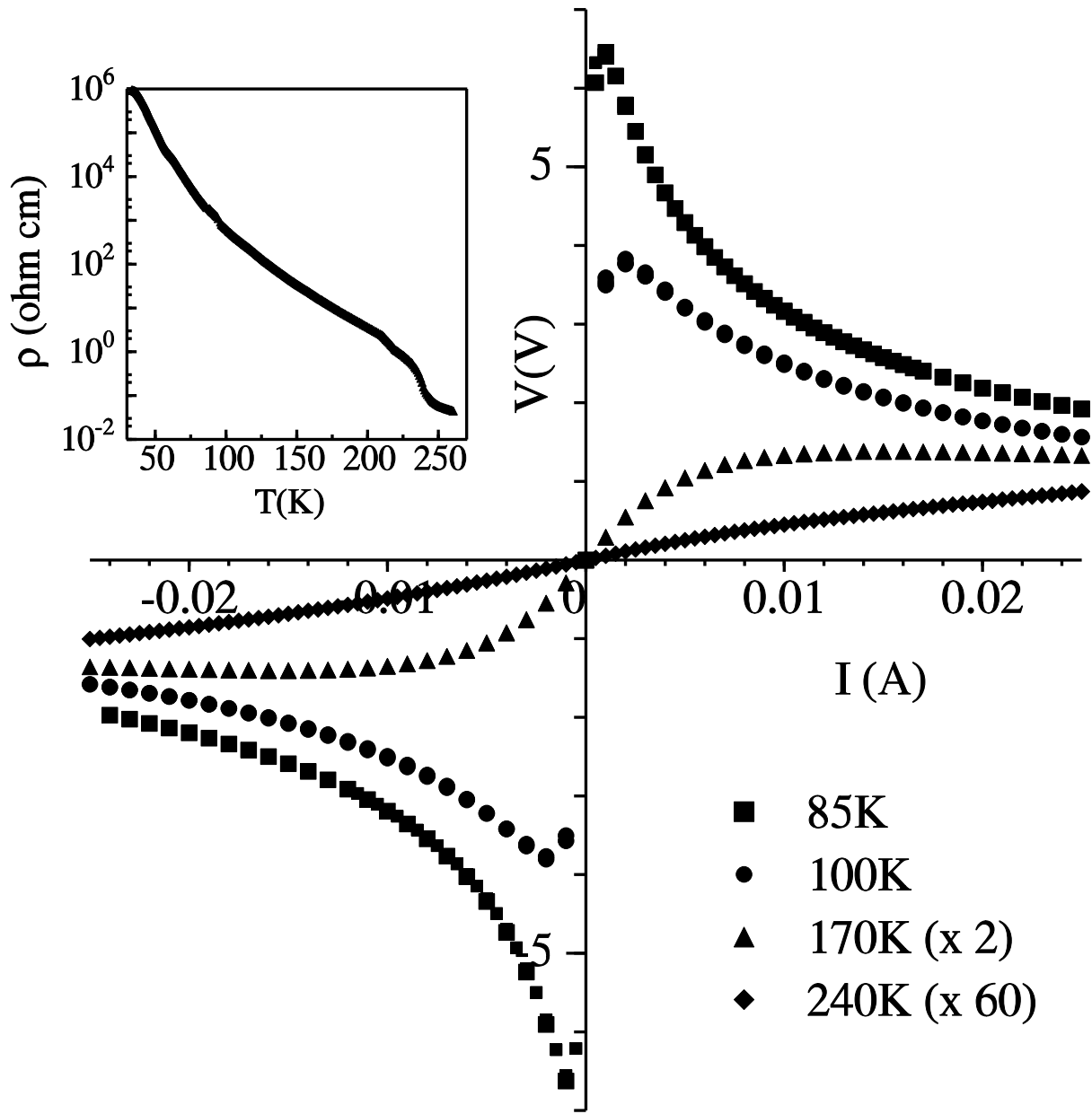
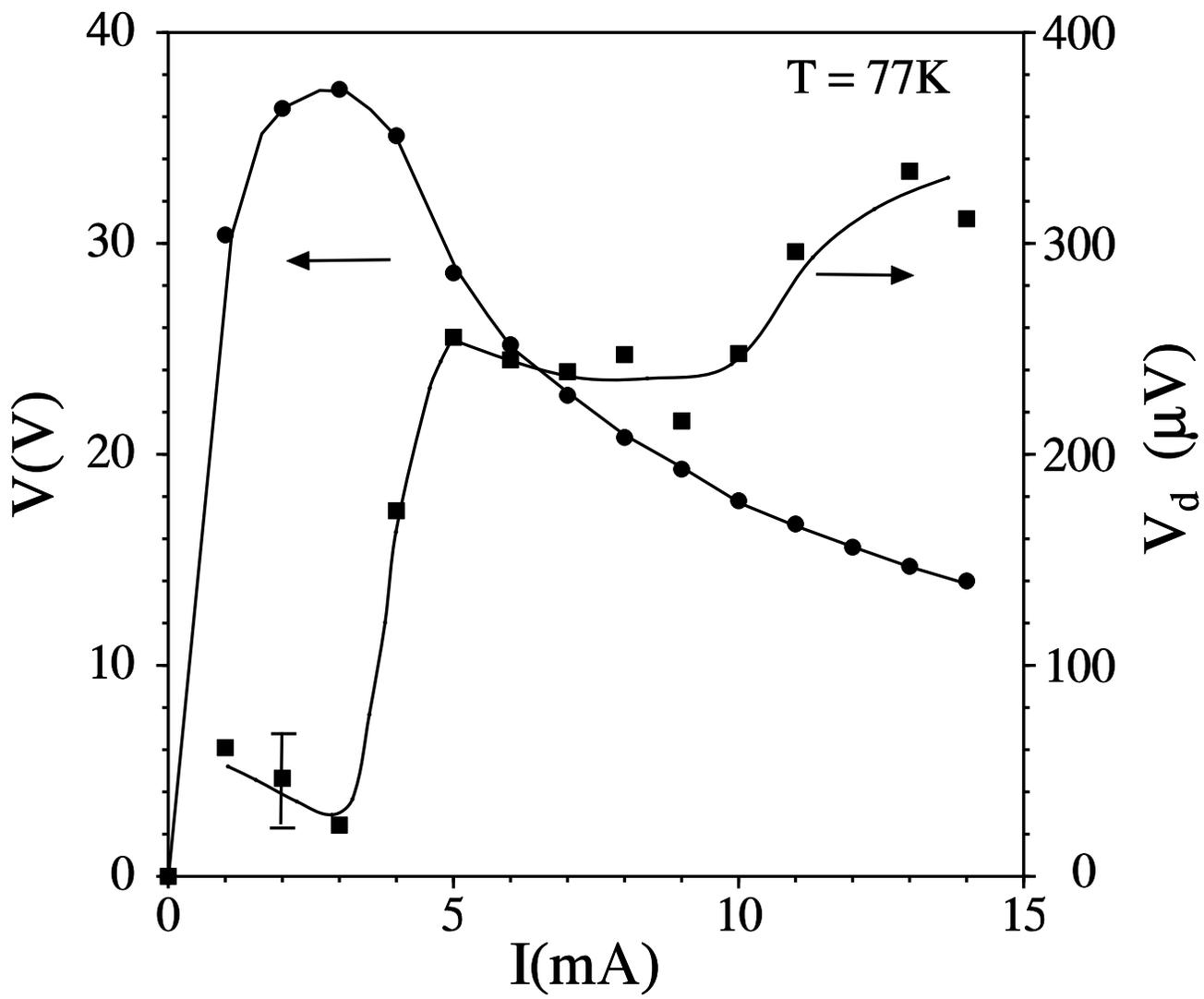
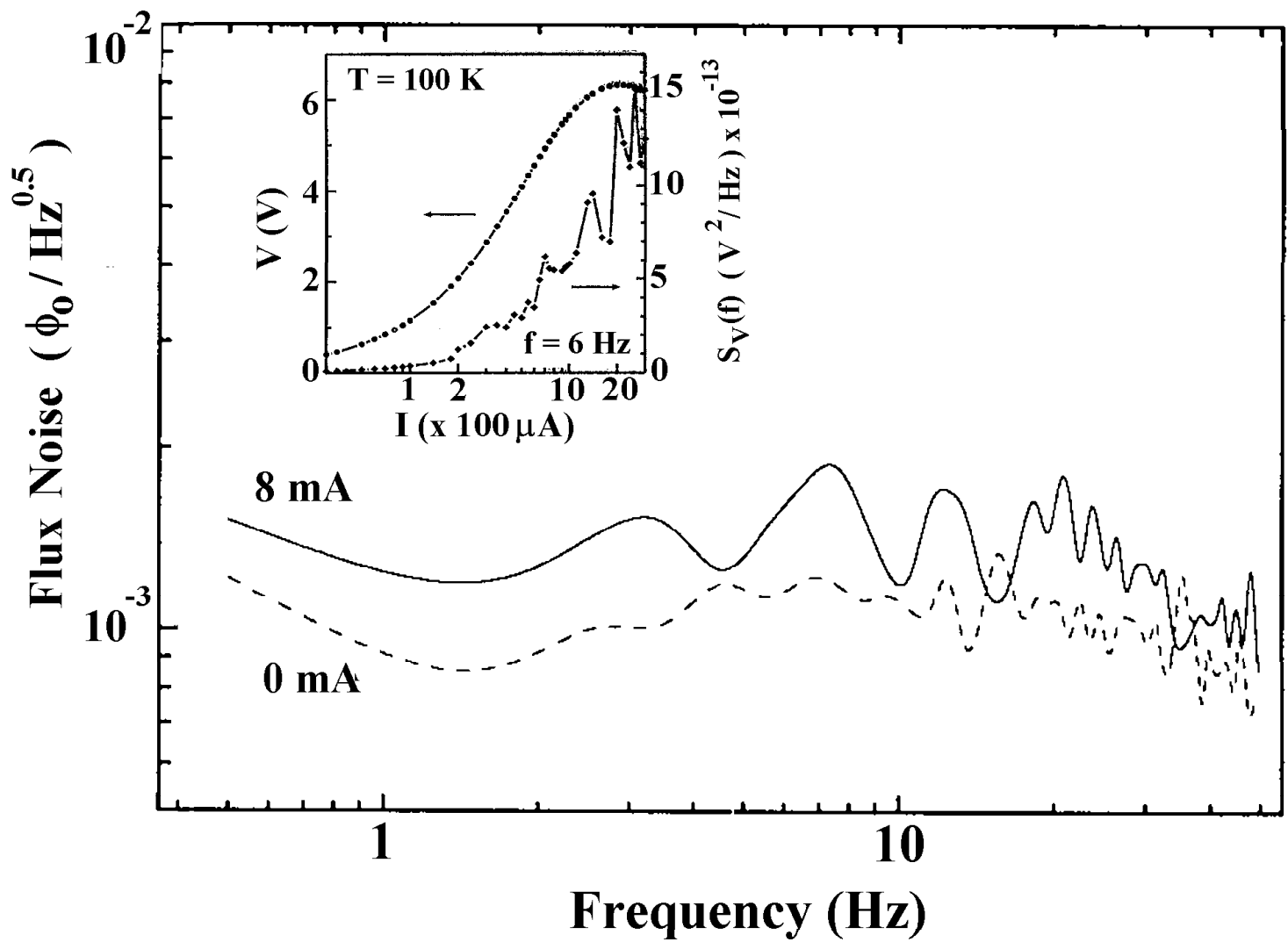
\end{document}